\documentclass{WileyMSP-template}
\usepackage{multicol}
\usepackage{braket}
\usepackage{float}
\usepackage{graphicx}
\usepackage{dcolumn}
\usepackage{bm}
\usepackage{amsmath}
\usepackage{cite}
\usepackage{xcolor}
\usepackage{amssymb}
\usepackage{gensymb}

\begin{document}

\rhead{\includegraphics[width=2.5cm]}

\title{Beamsplitter-free, high bit-rate, quantum random number generator based on temporal and spatial correlations of heralded single-photons}

\maketitle

\vspace{12pt}
\author{Ayan Kumar Nai},
\author{Amritash Sharma},
\author{Vimlesh Kumar},
\author{Sandeep Singh},
\author{Shreya Mishra},
\author{C. M. Chandrashekar},
\author{G. K. Samanta}





\vspace{12pt}
\begin{affiliations}
Ayan Kumar Nai\\
Photonic Sciences Lab., Physical Research Laboratory, Ahmedabad 380009, Gujarat, India\\
Indian Institute of Technology-Gandhinagar, Ahmedabad 382424, Gujarat, India\\
Email Address: ayankrnai@gmail.com\\
\vspace{12pt}

Amritash Sharma, Vimlesh Kumar, Sandeep Singh, Shreya Mishra, G. K. Samanta\\
Photonic Sciences Lab., Physical Research Laboratory, Ahmedabad 380009, Gujarat, India\\
\vspace{12pt}
C. M. Chandrashekar\\
Quantum Optics $\&$ Quantum Information, Department of Instrumentation and Applied
Physics, Indian Institute of Science, Bengaluru 560012, India
The Institute of Mathematical Sciences, C. I. T. Campus, Taramani, Chennai 600113, India

\end{affiliations}


\keywords{QRNG, heralded single-photon source, Second order-correlation, HOM interferometry, Entropy, Toeplitz, NIST, Auto-correlation}

\justifying
\begin{abstract}
The spontaneous parametric down-conversion (SPDC), an inherently random quantum process, produces non-deterministic photon-pair with strong temporal and spatial correlations owing to both energy and momentum conservation. Therefore, the SPDC-based photon pairs are used for quantum random number generation (QRNG). Typically, temporal correlation in association with an ideal unbiased beam splitter is used for QRNG without fully exploring the spatial correction. As a result, SPDC-based QRNG has a low bit rate. On the other hand, due to the spatial correlation, the photon pairs in non-collinear phase-matched geometry are generated randomly in diametrically opposite points over an annular ring spatial distribution. Therefore, exploring the temporal correlation between photon pairs from different sections of the annual ring can lead to multi-bit QRNG at a high rate, avoiding the need for a beam splitter. As a proof-of-concept, we report on high bit-rate QRNG by using spatial correlation of photon-pairs by sectioning the SPDC ring of a non-collinear, degenerate, high-brightness source and temporal correlation between the diametrically opposite sections. Dividing the annular ring of the high-brightness photon-pair source based on a 20 mm long, type-0 phase-matched, periodically-poled KTP crystal into four sections, recording the timestamp of the coincidences (widow of 1 ns) between photons from diametrically opposite sections and assigning bits (0 and 1), we extracted 90 million raw bits over 27.7 s at a pump power of 17 mW. We determined the extraction ratio using the minimum entropy evaluation and obtained more than 95$\%$ extraction of raw bits. Using Toeplitz matrix-based post-processing, we achieved a QRNG with a bit-rate of 3 Mbps, passing all NIST 800-22 and TestU01 test suites.  The generic scheme shows the possibility of further enhancement of bit rate through the use of more sectioning of the SPDC ring.


\end{abstract}

\begin{multicols}{2}
\section{Introduction}
\justifying
Random numbers have attracted a great deal of attention in many fields, such as fundamental physics tests, scientific simulations, numerical analysis, decision-making, cryptography, and many aspects of our everyday lives \cite{Miguell2017}. 
Traditional methods to generate random numbers often rely on mathematical algorithms. Despite the increased complexity, the algorithm-based random numbers have pseudorandomness as they can be predicted or replicated, compromising the security of systems \cite{James2020}. Therefore, to achieve true randomness - as essential for the highest levels of security, one needs to rely on random physical phenomena. Quantum mechanical processes \cite{Sch}, with their inherent probabilistic behavior, offer a natural and reliable basis for quantum random number generation (QRNG) with high security. \\
\indent Over the years, QRNGs have been explored in numerous approaches, including single-photon emitters \cite{Luo}, defect centers \cite{Chen}, phase noise \cite{Qi}, and quantum vacuum and phase fluctuations \cite{Hayl, Xu:12}. The truly random bits have also been explored using spontaneous parametric down-conversion (SPDC) based heralded single photons, and single photon entanglement states \cite{Leone, Xu:16}. One of the simplest implementations of a QRNG is the use of a single-photon source incident on a lossless 50:50 beam splitter and exploring the temporal correlation (antibunching) of the photons to generate random bits. However, these approaches require both an efficient single-photon source and a perfectly unbiased, lossless beam splitter, which are technologically challenging. The use of attenuated optical pulses has also been explored as an alternative to single-photon sources in these beam splitter-based QRNGs. Nevertheless, enhancing source efficiency, collection, detection efficiency, and addressing the bias in beam splitters remain critical challenges in QRNG implementation. Again, the low collection efficiency (ratio of coincidence counts and singles count) \cite{Jabir2017} of the heralded photons results in a lower bit rate for the QRNG system \cite{Leone}. Therefore, it is imperative to explore new experimental schemes to improve the overall bit rate of the QRNG systems based on the temporal correlation of heralded single photons. \\
\indent Here, we present a novel beam-splitter bias-free experimental scheme for a QRNG system with high bit rates based on the strong temporal and spatial correlations of photon pairs generated through the SPDC process. Using a 20-mm long periodically-poled potassium titanyl phosphate (PPKTP) crystal, pumped with a continuous-wave diode laser, we have generated non-collinear, degenerate SPDC photons in an annular intensity profile. Owing to the momentum matching condition, photon pairs are generated at diametrically opposite points with strong temporal correlations and exhibit random spatial distribution around the ring. By dividing the annular spatial profile into four sections and measuring coincidences between photons from opposite sections, we generate random bits with a min-entropy exceeding 95$\%$, ensuring high efficiency in extracting true random bits. Using Toeplitz matrix-based post-processing, our QRNG system achieves a bit rate of 3 Mbps, passing all NIST 800-22 and TestU01 statistical test suites. This experiment shows that the QRNG bit rate can be further enhanced by dividing the ring into additional sections, such as 8, 16, or 32, to develop a multi-bit QRNG system. 

\section{Experiment}
\begin{figure}[H]
    \centering
    \includegraphics[width=\linewidth]{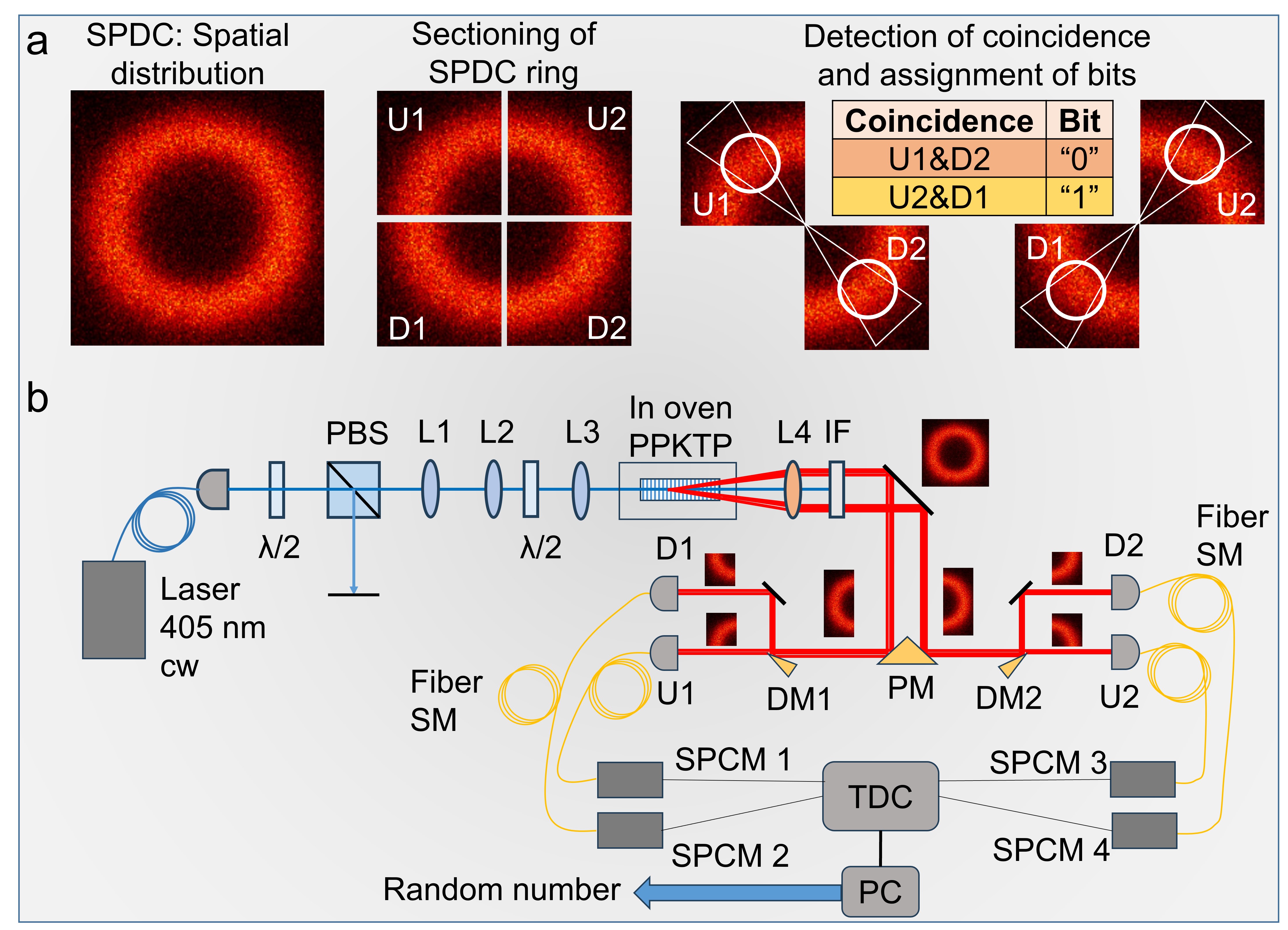}
    \caption{The schematic representation of (a) the concept and (b) the experimental setup of the QRNG source. Laser: 405 nm, cw diode laser, $\lambda$/2: Half wave plate, L1-4: plano-convex lenses of different focal lengths, PPKTP: 20 mm long PPKPT crystal of single grating period in temperature oven, IF: Interference filter, PM: Prism shaped gold coated mirror, DM1-2: D-shaped mirrors, U1-2 and D1-2: Upper and lower sections of the SPDC ring, SM: Single-mode fiber, SPCM1-4: single photon counting modules, TDC: Time-to-digital converter, PC: Computer. }
    \label{Figure1}
\end{figure}

The concept of QRNG using both temporal and spatial correlations, along with a schematic of the experimental setup, is illustrated in Fig. \ref{Figure1}. As shown in Fig. \ref{Figure1}(a), the annular spatial distribution of SPDC photons generated through degenerate, type-0, non-collinear phase-matching is divided into four sections labeled as upper (U1, U2) and lower (D1, D2). Since the diametrically opposite sections contain correlated photon pairs (spatial correlation), we measured coincidences (temporal correlation) between sections (U1, D2) and (U2, D1), subsequently assigning the bits "0" and "1," respectively. As the photon pairs randomly appear in sections (U1, D2) and (U2, D1), this method produces a sequence of random bits of 0 and 1. The novel concept is implemented using the schematic shown in Fig. \ref{Figure1}(b), where a continuous-wave, single-frequency laser diode generates an output power of 25 mW at a central wavelength of 405 nm (Integrated Optics MatchBox 0405L-25A), with a spectral bandwidth (full width at half maximum) of approximately 20 MHz, acting as the pump laser. Operating the laser at its highest power to ensure optimum performance, the power attenuator composed of a half-wave plate ($\lambda/2$ plate) and a polarizing beam splitter (PBS) cube is used to control the laser power in the experiment. A second $\lambda/2$ plate, placed after the beam collimating lenses, $L_1$ (focal length, $f$ = 100 mm) and $L_2$ (focal length, $f$ = 150 mm), is used to adjust the polarization state of the pump laser to ensure optimum phase-matching, depending on the orientation of the nonlinear crystal. A 20-mm long, 1$\times$2 mm$^2$ aperture periodically-poled potassium titanyl phosphate (PPKTP) crystal, with a single grating period of $\Lambda$ = 3.425 $\mu$m, is used to generate degenerate SPDC photons at 810 nm through non-collinear, type-0 (e $\rightarrow$ e + e) phase-matching. The crystal is housed in an oven, with the temperature variable up to 200$\degree$C and stability of $\pm$0.1$\degree$C. The SPDC photons generated with the pump beam focused at the center of the PPKTP crystal by a lens, $L_3$ (focal length, $f$ = 150 mm with a spot size of $\sim$40 $\mu$m, are collimated using the lens, $L_4$, of focal length, $f$ = 100 mm. The collimated annular SPDC ring, as observed by the EMCCD camera (Andor, iXon Ultra 897) after extraction from the pump by the interference filter (IF) of bandwidth $\sim$10 nm centered at 810 nm, is first divided into two half-circles using the gold-coated prism mirror (PM). Subsequently, the half-circles of the SPDC ring are divided into four quarters, named U1, D1, U2, and D2, using the D-shaped mirrors, DM1 and DM2, and collected using the combination of a fiber coupler, single-mode fiber (SM) and single photon counting modules (SPCM-AQRH-14-FC), SPCM1-4. The time-to-digital converter (TDC) (MultiHarp 150, PicoQuant) records the timestamp of the coincidence events of the photons collected from diametrically opposite quarters of the SDPC ring and produces a sequence of random bits of 0's and 1's. The selection of a coincidence window plays a vital role in detecting true coincidences while avoiding false positives. For example, a too-narrow coincidence window can miss true coincidences, and a too-wide window can include uncorrelated events, thus underestimating and overestimating the experimental parameters. Since we are dealing with a high-brightness single photon source \cite{Jabir2017, singh2023AQT,  Singh23b} in the current experiment, keeping pump power of 1 mW, we have measured $g^{(2)}(\tau)$ with the coincidence window width in a step of 0.5 ns starting from 0.5 ns and found the increase of $g^{(2)}(\tau)$ from 0.016 to 0.1 at a slope of 0.019 per nanosecond width of the coincidence window. Since the timing jitter of the SPCM and electronics can collectively be estimated to be in the range of 0.6 ns, we found the coincidence window of 1 ns is optimum to handle our high brightness source without significantly changing the overall coincidence counts. Therefore, we have set the coincidence window of 1 ns, (200 bins of width 5 ps) throughout the studies unless otherwise mentioned. 
\section{Results and discussions}
\begin{figure}[H]
    \centering
    \includegraphics[width=\linewidth]{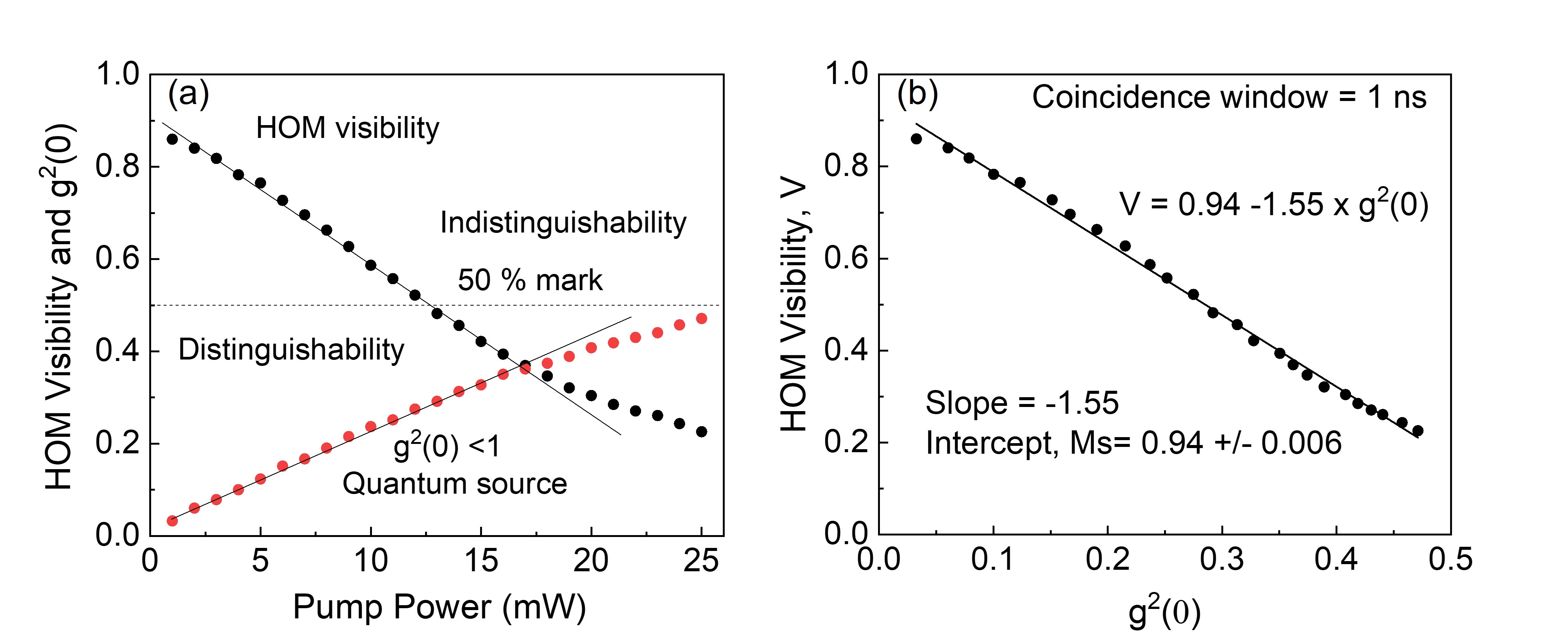}
    \caption{(a) Variation of HOM visibility (indistinguishability), and $g^{(2)}(0)$ (purity) of the heralded single photons as a function of pump power. (b) Dependence of HOM visibility on change of $g^{(2)}(0)$. }
    \label{Figure2}
\end{figure}

We first characterized the photon pair source purity and indistinguishability as a function of pump power using Hong-Ou-Mandel (HOM) interferometry \cite{Hong} and $g^{(2)}(\tau)$ \cite{Paper14}. While these two experiments can simultaneously be done using the photon pairs of diametric opposite sections (see Fig. \ref{Figure1}(b)), we used photon pairs from sections U1 and D2 and performed two measurements one after the other. The results are shown in Fig. \ref{Figure2}. As expected, it is evident from Fig. \ref{Figure2}(a) that the HOM visibility, V (black dots), decreases from 0.86 to 0.22, and the $g^{(2)}(0)$ (red dots) increases from 0.032 to 0.47 with an increase in pump power from 1 mW to 25 mW. However, we observe a deviation from the linear dependence of both HOM visibility, V, and $g^{(2)}(0)$, at higher pump power due to the saturation of the single photon detectors beyond single counts of  12 - 14 MHz, as obtained for the pump power in the range of 12 - 15 mW. It is interesting to note that although the source has a maximum $g^{(2)}(0)$ value of 0.47, which is well below the classical limit ($g^{(2)}(0)$$<$1) find appropriate citation\cite{Kimble}, ensuring the non-classical nature of the light sources within the coincidence window of 1 ns, the lower HOM visibility (V = 0.22) confirms the distinguishability of photons due to the generation of multiphotons at higher pump powers. Increasing the width of the coincidence window would escalate the rate of change in both HOM visibility and $g^{(2)}(0)$ with pump power.

\begin{figure}[H]
    \centering
    \includegraphics[width=0.8\linewidth]{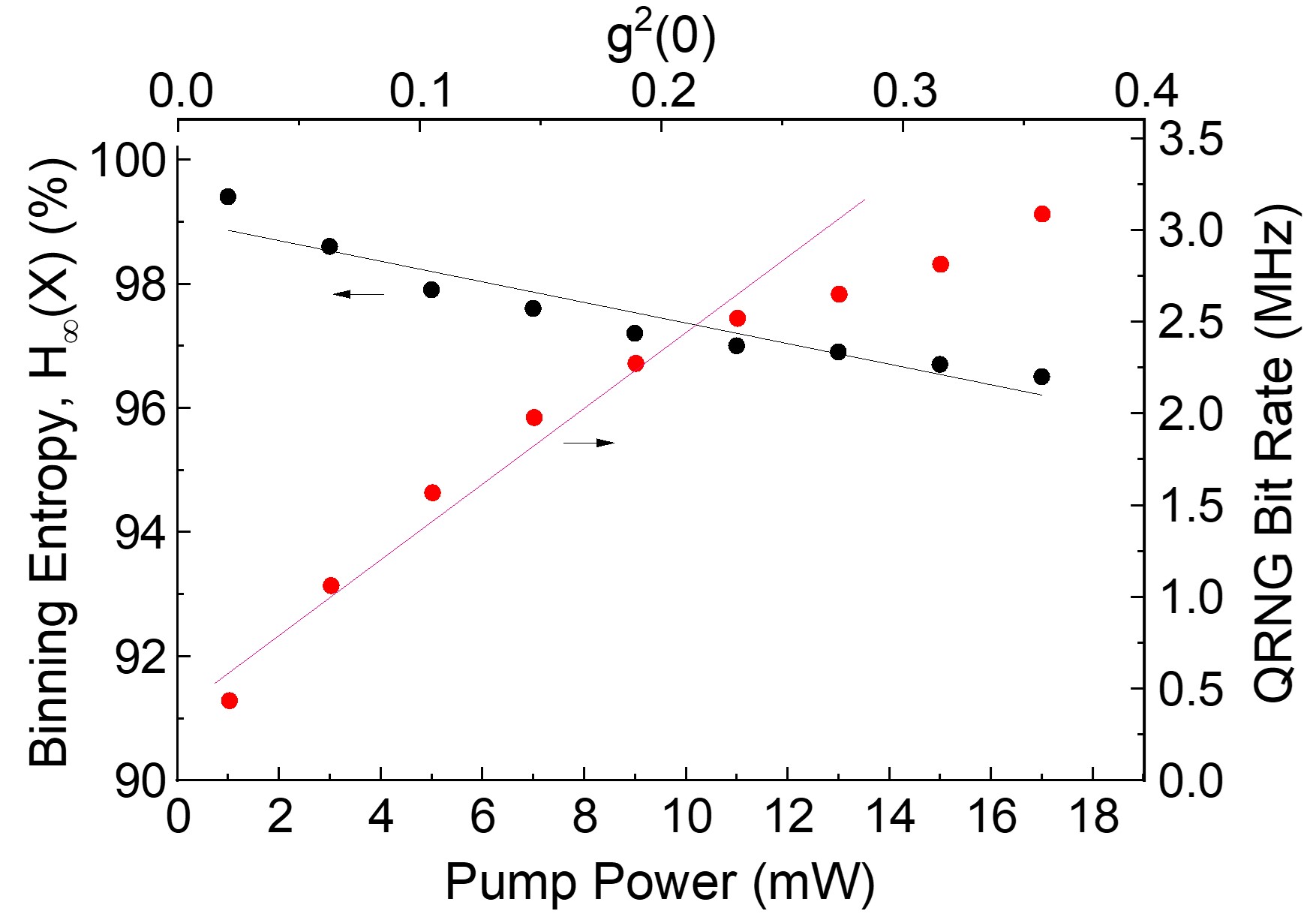}
    \caption{The variation of the minimum entropy, $H_{\infty}(X)$ (black dots), and post-processed QRNG bit rate (red dots) as function of pump power and $g^{(2)}(0)$. The solid lines (red and black) are linear lines fit to the experimental data.}
    \label{EntvsPower}
\end{figure}

To gain further insight into the single-photon characteristics and distinguishability between photons in the presence of multiphoton generation due to increased pump power, we plotted HOM visibility (V) as a function of $g^{(2)}(0)$ using the data from Fig. \ref{Figure2}(a). The results, shown in Fig. \ref{Figure2}(b), reveal that the HOM visibility and $g^{(2)}0)$ for a fixed coincidence window width (here, 1 ns) follow a linear relationship (black line) given by V = 0.94 - 1.55 $\times$ $g^{(2)}(0)$, different from the relationship V$\approx$ 1 - $g^{(2)}(0)$ predicted for a high purity single-photon source based on quantum dots \cite{Trivedi:20, Ollivier:21}. The intercept value (0.94 $\pm$ 0.006), which defines the intrinsic single-photon indistinguishability of the current source, is slightly lower due to the transmission function of the interference filter with a bandwidth of approximately 10 nm. Meanwhile, the slope of the curve indicates the generation and collection of multiphotons, resulting in photon distinguishability. Both the slope and the intercept value can change depending on the collection optics, such as the use of a multimode fiber. We observed similar performance of the source in terms of HOM visibility and $g^{(2)}(0)$ when using photon pairs collected from sections U2 and D1, confirming the uniform performance of the source throughout the annular distribution. These results suggest that one needs to selectively use either parameter to characterize the single-photon source based on the intended application. For example, for HOM interferometer-based quantum sensing \cite{singh2023AQT}, HOM visibility may be a more relevant measure of source quality than $g^{(2)}(0)$$<1$.\\ 
\indent Knowing the performance metric of the heralded single-photon source as a function of input power, we recorded the coincidence counts for the pair photons collected from the sections (U1 $\&$ D2) and (U2 $\&$ D1) (see Fig. \ref{Figure1}) and assigned 0's and 1's, respectively. Subsequently, we have recorded more than 90 millions of raw bits of 0's and 1's for each pump power requiring varying bit recording time, higher power smaller time and lower pump power, higher time due to the lower photon generation rate. Since the raw bit contains all possible errors, we calculated the minimum entropy \cite{Xu:12, Ma2013} to determine the maximum extraction of final, unbiased, true random bits. The min-entropy of a probability distribution $X$ on $\{0,1\}^N$ is defined by,
\begin{equation}
H_{\infty}(X)=-\log _2\left(\max _{x \in\{0,1\}^N} P_r[X=x]\right)
\label{Eq3}
\end{equation}
To calculate the min-entropy $H_{\infty}(X)$, we applied an 8-bit ($N = 8$) binning technique by segmenting the bit string into 8-bit blocks, each mapped to one of the 256 possible bins ($2^8$). This resulted in 256 sample points ($\{0,1\}^N$), each representing a unique 8-bit sequence. We identified the most probable sample, denoted as $x$, from this set and calculated the min-entropy using the standard formula as represented by Eq. \ref{Eq3}. 
As seen in Fig.~\ref{EntvsPower}, the min-entropy, $H_{\infty}(X)$ (black dots), decreased from 99.4\% to 96.5\% as the pump power increased from 1 mW to 17 mW, and subsequent increase of $g^{(2)}(0)$ from 0.03 to 0.36 due to the rise in multi-photon events at higher pump powers. Despite this, the raw bit string retained more than 0.96 bits per bit of randomness, enabling the extraction of over 96\% quantum random bits. However, the high $H_{\infty}(X)$ value of 99.4\% at 1 mW indicates the potential for generating pure quantum random numbers at a lower bit rate, highlighting a trade-off between min-entropy and bit rate.
 To distill true random bits from the raw bit sequence, we employed a Toeplitz hashing randomness extractor \cite{krawczy}. We constructed a binary Toeplitz matrix of size $m \times n$  using a pseudorandom seed of $m$ + $n$ - 1 = 175.5 million bits, with $m = 85.5$ million bits, 95\% of the raw bits, $n = 90$ million bits ($m$ = 0.95 $\times$ $n$). To overcome the computational limitations, we divided the raw and seed sequences into blocks of one million bits, post-processed each block separately, and concatenated all the bits to make the string of final post-processed bits. The red dots in Fig.~\ref{EntvsPower} show that the post-processed QRNG bit rate increased linearly with pump power, from 0.4 million bits per second at 1 mW to 2.5 million bits per second at 11 mW, with a slope of 0.2 million bits per second per mW of pump power. Beyond 11 mW, the bit rate deviated from linearity due to detector saturation, reaching a maximum of 3.08 million bits per second at 17 mW. Further improvement in the QRNG bit rate can be possible using higher pump power in combination with the use of a detector with high quantum efficiency and a high saturation threshold, such as superconducting nanowire single-photon detectors (SNSPDs) and a lower coincidence widow ($<$ 1 ns used in the current study).

\begin{figure}[H]
    \centering
    \includegraphics[width=\linewidth]{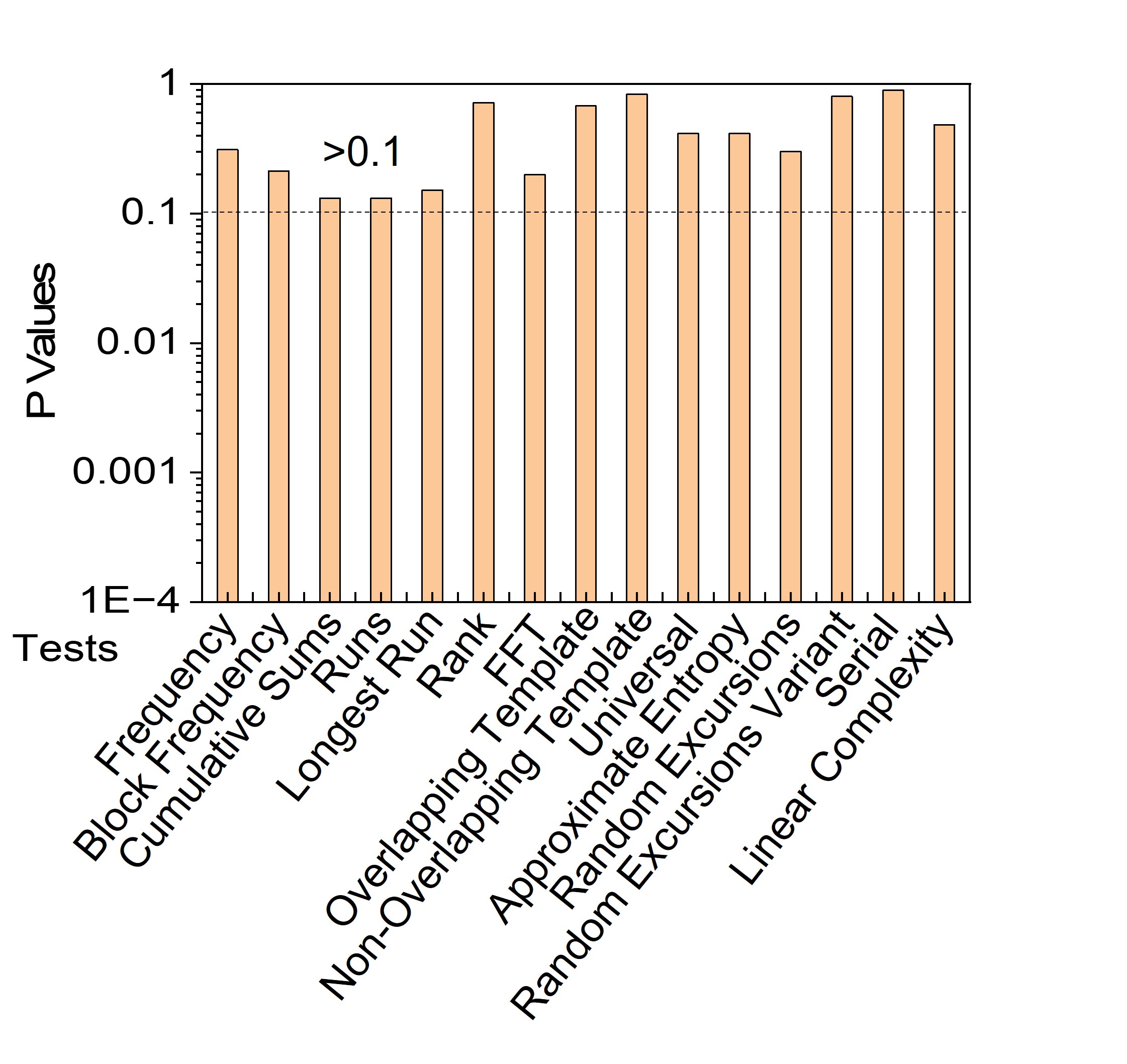}
    \caption{Final P-values of 15 tests under NIST statistical suit for the Toeplitz post-processed bits recorded for a pump power of 17 mW.}
    \label{NIST}
\end{figure}

After having the post-processed QRNG bits, we used the NIST 800-22 test suite \cite{rukhin} to evaluate the quality of randomness of the post-processed bits. We have divided the post-processed bit string into 80 sequences (higher than the required minimum samples, 55 sequences of NIST test suite) of binary bits, each containing one million bits as the input file for the NIST test suite. As the NIST statistical test suite follows two primary types of quantification to determine the randomness of the sequence, we calculated the uniformity of the P-values and the proportionality test to find the proportion of the input sequences passing (P-value is above the chosen significance level, $\alpha$, usually $\alpha$ = 0.01) a test.
Using the Goodness-of-Fit distributional test and Kolmogorov-Smirnov tests to the post-processed bits, we found the final P-values of 15 different tests under the NIST test suite of the post-processed bits for a pump power of 17 mW are shown in Fig. \ref{NIST}. As evident from Fig. \ref{NIST}, the P-values of all the tests are $>$ 0.1, confirming the randomness of the experimental results. 

Again, we recorded the proportionality test results for the post-processed bit sequences in Table \ref{table1}. According to the NIST test suite, the proportional fraction should lie within the range given by the equation,
\begin{equation}
(1-\alpha) \pm 3 \sqrt{\frac{\alpha(1-\alpha)}{n}}
\label{Eq2}
\end{equation}
where $n$ is the number of sequences. For 80 sequences, each consisting of 1 million bits, NIST considers $n = 49$ for the random excursions and random excursion variant tests. Using $\alpha = 0.01$ and $n = 49$ in Eq. \ref{Eq2}, the calculated range for the proportional fraction is (0.947, 1.033). For the remaining tests, using $\alpha = 0.01$ and $n = 80$, the proportional fraction range is (0.956, 1.023). As shown in Table \ref{table1}, the proportional fractions for all post-processed bit sequences lie within the defined ranges, indicating successful results across all tests.


\begin{table}[H]
\caption{Results of the NIST statistical tests}
\label{table1}
\centering
\small
\begin{tabular}{lrl}
\hline
\hline
Test name & Proportion of & Result \\
& successful sequences & \\

\hline
Approximate entropy           & 0.988                             & Success \\
Frequency within a block      & 0.975                              & Success \\
Cumulative sums               & 0.975                              & Success \\
Discrete Fourier transform    & 0.988                             & Success \\
Frequency                     & 0.975                              & Success \\
Longest run in a block        & 0.962                              & Success \\
Non-overlapping template      & \(\min 0.962\)                     & Success \\
                              & \(\max 1.000\)                     & Success \\
Overlapping template          & 1.000                              & Success \\
Random excursions             & \(\min 0.980\)                     & Success \\
                              & \(\max 1.000\)                     & Success \\
Random excursions variant     & \(\min 0.960\)                     & Success \\
                              & \(\max 1.000\)                     & Success \\
Binary matrix rank            & 1.000                              & Success \\
Runs                          & 0.988                              & Success \\
Serial                        & 0.975                              & Success \\
Linear complexity             & 0.988                              & Success \\
Maurer's universal            & 0.988                              & Success \\
\hline
\hline
\end{tabular}
\end{table}

Further, we evaluated the randomness of our post-processed bits using the TestU01 statistical test suite, specifically the Rabbit, Alphabit, and BlockAlphabit battery tests. While these tests typically require sequences of approximately 30 million bits, we employed longer sequences of 80 million bits for more rigorous analysis. Following the methodology outlined in the literature \cite{testu01, Neil}, we interpreted the P-values by considering values
within the interval $\left[10^{-3}, 1-10^{-3}\right]$  as a success, and value outside this interval as a failure. The results of the TestU01 tests on our post-processed random bits, as shown in Table \ref{table2}, indicate that our novel experimental scheme successfully passed all the tests, regardless of the interpretation method applied.

\begin{table}[H]
\caption{Results for the TestU01 statistical tests }
\label{table2}
\centering
\small
\begin{tabular}{lrl}
\hline
\hline

Battery & Sequence length & Result \\
\hline \\
Rabbit & $8 \cdot 10^7$ & Success \\

Alphabit & $8 \cdot 10^7$ & Success \\

BlockAlphabit & $8 \cdot 10^7$ & Success \\
\hline
\hline
\end{tabular}
\end{table}

\begin{figure}[H]
    \centering
    \includegraphics[width=\linewidth]{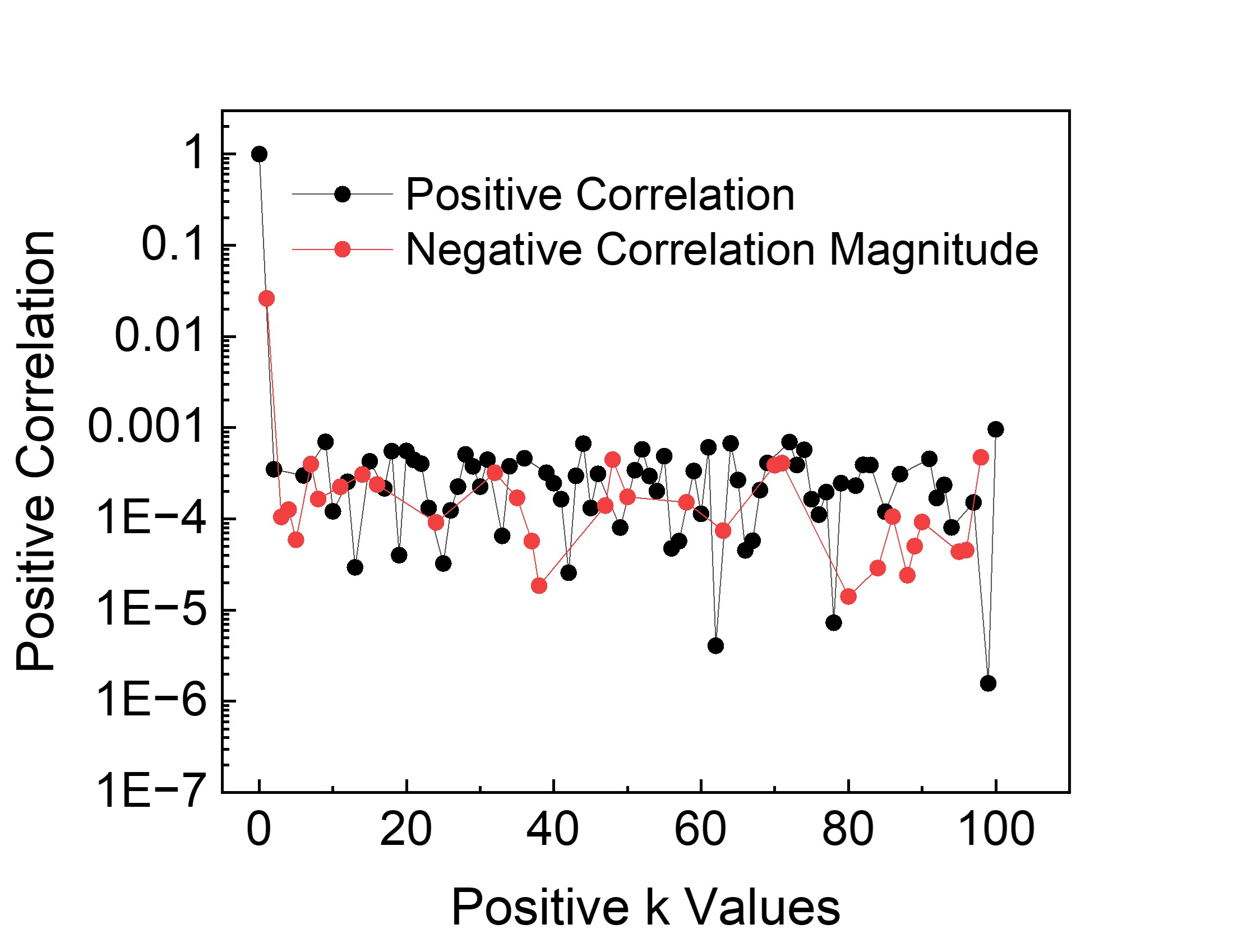}
    \caption{Variation of the magnitude of the positive and negative autocorrelation coefficients of the first 10 million bits of the raw bit sequence up to a delay of 100 bits. The raw bit sequence is recorded for the pump power of 17 mW.}.
    \label{Auto}
\end{figure}

We also analyzed the randomness of the raw data recorded for the pump power of 17 mW by calculating the autocorrelation coefficient. Using the first 10 million bits of the raw bit sequence up to a delay of 100 bits, we computed the autocorrelation coefficient, with the results shown in Fig. \ref{Auto}. As evident from the figure, even after a delay of 100 bits, the mean and standard deviation of the autocorrelation coefficient do not change significantly. Compared to a truly random sequence, which would have a minimum autocorrelation coefficient of zero mean value, the raw bits from our experiment exhibit a minimum autocorrelation coefficient with a mean value on the order of \(10^{-6}\) and a standard deviation of \(8.326 \times 10^{-6}\). Such low values for the minimum autocorrelation coefficient confirm that the raw bit string recorded from the current experiment represents a good random sequence, even prior to post-processing. This observation further certifies the high value of the min-entropy, \(H_{\infty}(X)\), and extraction efficiency.

\section{Conclusions}
In conclusion, we presented a novel QRNG scheme based on the temporal and spatial correlations of heralded single-photon sources, bypassing the need for traditional beam splitters and enhancing both the efficiency and simplicity of the overall system architecture. We characterized the purity and indistinguishability of the single-photon sources using the second-order correlation function $g^{(2)}(0)$ and Hong-Ou-Mandel (HOM) visibility, finding a linear relation between these two parameters for the heralded single-photon sources. Interestingly, while the purity of the single-photon source, quantified by the $g^{(2)}(0)$ value, determines the min-entropy for the maximum extractable true random bits from the raw bit sequence, the photon indistinguishability, indicated by HOM visibility, shows weak or no dependence on QRNG performance.

For example, at a pump power of 17 mW, the heralded single-photon source demonstrated $g^{(2)}(0) = 0.36$ and HOM visibility of 36$\%$. Nonetheless, the min-entropy of the raw bit sequence was found to be 0.96, with a very low autocorrelation coefficient ($\sim 10^{-6}$) even after a delay of 100 bits. This suggests that QRNG can be developed without the need for coincidence detection between the photons \cite{shafi2023multi}. After applying Toeplitz post-processing to the raw bit sequence, we confirmed that the output passed both the NIST and TestU01 test suites, verifying the development of a high-quality QRNG with a high bit rate. Further enhancement in the bit rate from a few hundred MHz to GHz can be made possible using multi-bit QRNG  by multiple sections of the SPDC ring or by recording the photon pair events using a single-photon avalanche diode (SPAD) camera and assigning the bits for the coincidence between the diametrically oppositive pixels. Our QRNG scheme offers a promising path to achieve a device-independent QRNG source with higher bit rates while ensuring true randomness, making it highly suitable for secure applications across various fields. 

\section*{ACKNOWLEDGMENTS}
A. K. N., A. S., V. K., S. S., S. M., and G. K. S. acknowledge the support of the Department of Space, Govt. of India. A. K. N. acknowledges funding support for Chanakya - PhD fellowship from the National Mission on Interdisciplinary Cyber-Physical Systems of the Department of Science and Technology, Govt. of India through the I-HUB Quantum Technology Foundation.  G. K. S. acknowledges the support of the Department of Science and Technology, Govt. of India, through the Technology Development Program (Project DST/TDT/TDP-03/2022).

\section*{AUTHOR DECLARATIONS}
\subsection*{Conflict of Interest}
The authors have no conflicts to disclose.
\subsection*{Author Contributions}
A. K. N. developed the experimental setup and performed measurements. A. S., V. K., S. S., and S. M. participated in experiments, data analysis, and numerical simulation with A.K.N. C. M. C. participated in the analysis and data interpretation. G. K. S. developed the ideas and led the project. All authors participated in the discussion and contributed to the manuscript writing.

\section*{DATA AVAILABILITY}
The data that support the findings of this study are available from the corresponding author upon reasonable request.

\bibliographystyle{IEEEtran}
\bibliography{Bib}


\end{multicols}
\end{document}